\documentclass[journal]{IEEEtran}

\usepackage{cite}

\ifCLASSINFOpdf
   \usepackage[pdftex]{graphicx}
\else
 
   \usepackage[dvips]{graphicx}
\fi

\usepackage[cmex10]{amsmath}
\usepackage{array}

\usepackage{url}

\usepackage{amsfonts}
\usepackage{amsmath}
\usepackage{amssymb}
\usepackage{amsbsy}

\usepackage{mathrsfs}

\usepackage{latexsym,bbm}
\usepackage{etex}
\usepackage{pstricks} 
\usepackage{tikz}
\usetikzlibrary{matrix,arrows}

\begin{document}

\title{Simulation of Ultra-thin Solar Cells Beyond the Limits of the Semi-classical Bulk Picture}

\author{Urs Aeberhard
\thanks{Urs Aeberhard is with the Institute of Energy and Climate Research - Photovoltaics (IEK-5), Forschungszentrum J\"ulich, 52425 J\"ulich, Germany
e-mail: u.aeberhard@fz-juelich.de.}}


\maketitle

\begin{abstract}
The photovoltaic characteristics of an ultrathin GaAs solar cell with a gold back reflector are simulated using the standard semiclassical drift-diffusion-Poisson model and an advanced microscopic quantum-kinetic approach based on the nonequilibrium Green's function (NEGF) formalism. For the standard assumption of flat-band bulk absorption coefficient used in the semiclassical model, substantial qualitative and quantitative discrepancies are identified between the results of the two approaches. The agreement is improved by consideration of field-dependent absorption and emission coefficients in the semiclassical model, revealing the strong impact of the large built-in potential gradients in ultrathin device architectures based on high-quality crystalline materials. The full quantum-kinetic simulation results for the device characteristics can be reproduced by using the NEGF generation and recombination rates in the semiclassical model, pointing at an essentially bulk-like transport mechanism.
\end{abstract}


\section{Introduction}

\IEEEPARstart{S}{olar} cells based on III-V semiconductors as absorber materials have been used for a long time in situations where high efficiency is the central requirement, as in space applications. Beyond those niches, the need for expensive wafer-based or epitaxial growth methods and the limited abundance of the materials have thus far restricted economic viability and the widespread application of the technology. A new opportunity for III-V solar cells was created with the effort in high-efficiency thin-film architectures targeting light-weight wearable devices operating close to the single-junction Shockley-Queisser limit \cite{kayes:11,miller:12,liu:12,wang:13_jpv,steiner:13}. Recently, there has been growing activity related to the scaling of the absorber thickness down to ultrathin layers below 100 nm\cite{massiot:12,wang:13,massiot:14,yang:14,vandamme:15}. In these devices, low single pass absorption needs to be compensated by plasmonic or nanophotonic absorption enhancement exploiting strong field enhancement and coupling to guided modes. 

While optical simulations have been instrumental in the design and optimization of light-trapping and absorber structures, electrical simulations of these ultrathin photovoltaic devices have received little attention so far. This might be related to the apparent simplicity of the device structure, which, in the past, has never made it necessary to go beyond the conventional drift-diffusion model for bulk materials. However, it is not obvious that an absorber as thin as a few tens of nanometers and subject to correspondingly large built-in fields should still behave as homogeneous bulk. Furthermore, it is common to block minority carriers at the contacts by corresponding barrier layers, which results in regions where transport may no longer be described by semiclassical theory \cite{cavassilas:15}. 

In this paper, the validity of the semiclassical picture for the simulation of ultrathin GaAs devices is assessed by direct comparison with an advanced quantum-kinetic approach based on the nonequilibrium Green's function formalism \cite{ae:jcel_11}. In this comprehensive microscopic approach, most of the questionable approximations of the semiclassical theory such as, extended bulk states, thermalized distributions, and absence of nonlocal processes, are relaxed, which provides a sound basis for the critical assessment of like approximations.       

The paper is organized as follows. In Section \ref{sec:simmod}, the semi-classical and quantum-kinetic simulation approaches are outlined. Their respective implementations and parametrizations are reported in Sec.~\ref{sec:modimp}. Section \ref{sec:resdisc} provides the numerical simulation results and a discussion of the discrepancies. A brief summary and conclusion is given in Section \ref{sec:sumconc}.

\section{Simulation approaches\label{sec:simmod}}

\subsection{Model System}

\begin{figure}[t]
\begin{center}
\includegraphics[width=0.5\textwidth]{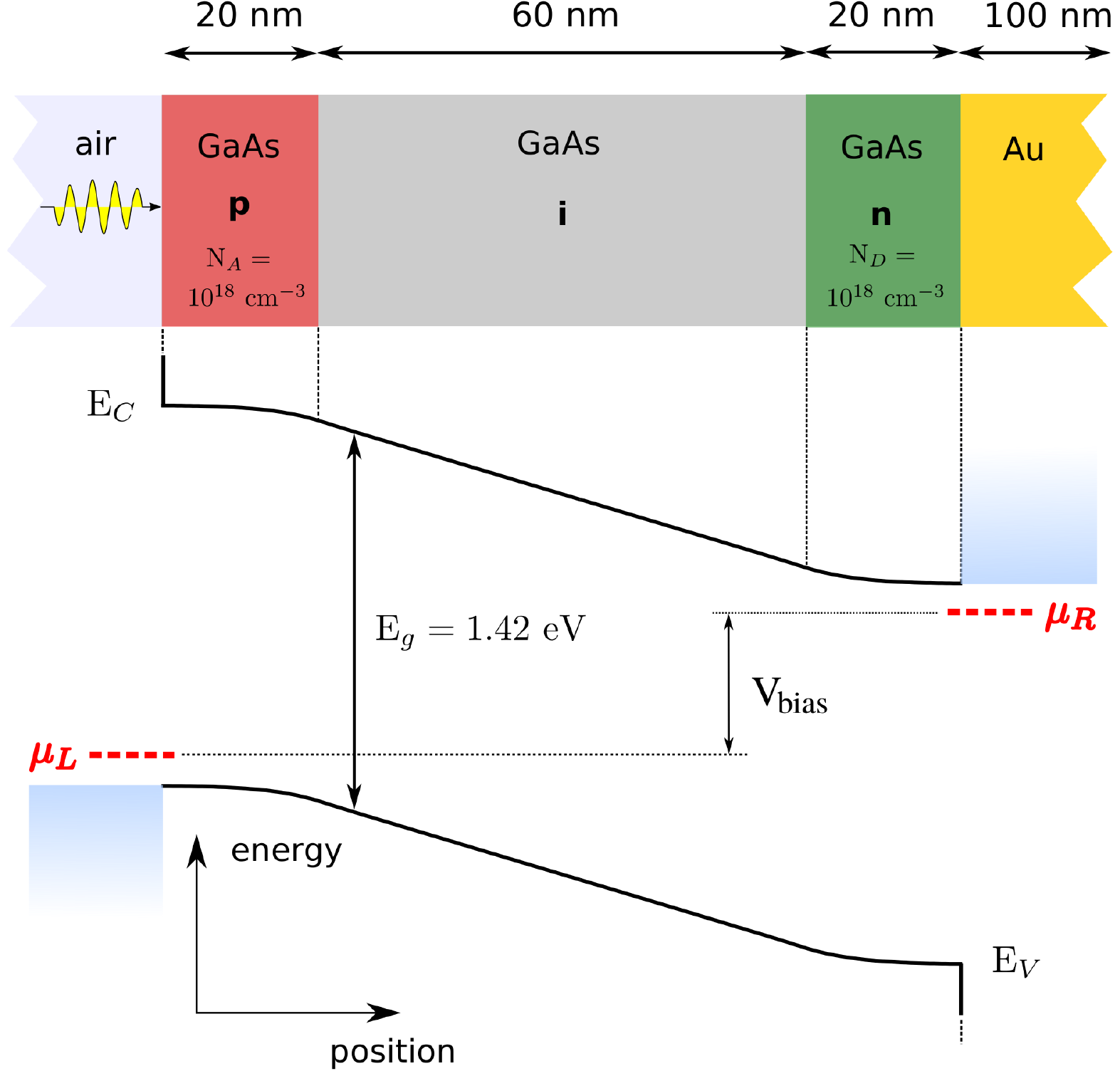}
\caption{Model system used in the simulations. The gold reflector is only used for the optical simulation. Ohmic contacts are assumed for majority carriers. \label{fig:structure_bands}}
\end{center}
\end{figure}

The model system used in the simulations is a GaAs \mbox{$p$-$i$-$n$} diode as shown in Fig.~\ref{fig:structure_bands}. The absorber consists of a 100-nm-thick GaAs slab with a 60-nm intrinsic region sandwiched between 20-nm-wide doped layers. The concentration of active dopants is $N_{A,D}=10^{18}$ cm$^{-3}$. Light is incident from the $p$-side. The gold reflector at the $n$-side is only used in the optical simulation. For majority carriers, ohmic contacts are assumed, i.e., there is no offset or Schottky barrier between the electrodes and the doped layers.

\subsection{Semiclassical Device Simulation: Drift-Diffusion-Poisson}

The conventional semiclassical approach to the simulation of $p$-$i$-$n$ solar cells at the steady state consists of the bipolar continuity equations for the charge densities $\rho_{n,p}$ ($n$: electrons, $p$: holes) under the assumption of a drift-diffusion current $\mathbf{j}_{n,p}$,
\begin{align}
-&\frac{1}{q}\nabla\cdot\mathbf{j}_{n}(\mathbf{r})=~\mathcal{G}(\mathbf{r})-\mathcal{R}(\mathbf{r}), \qquad (\textrm{electrons})\label{eq:conteq_el}\\
&\mathbf{j}_{n}(\mathbf{r})=q\left\{-\rho_{n}(\mathbf{r})\mu_{n}(\mathbf{r})\nabla
\phi(\mathbf{r})+D_{n}(\mathbf{r})\nabla\rho_{n}(\mathbf{r})\right\},\label{eq:ddcurr_el}\\
&\frac{1}{q}\nabla\cdot\mathbf{j}_{p}(\mathbf{r})=~\mathcal{G}(\mathbf{r})-\mathcal{R}(\mathbf{r}), \qquad (\textrm{holes})\label{eq:conteq_hl}\\
&\mathbf{j}_{p}(\mathbf{r})=-q\left\{\rho_{p}(\mathbf{r})\mu_{p}(\mathbf{r})\nabla
\phi(\mathbf{r})+D_{p}(\mathbf{r})\nabla\rho_{p}(\mathbf{r})\right\},\label{eq:ddcurr_hl}
\end{align}
coupled to the bipolar Poisson equation for the electrostatic (Hartree) potential $\phi$,
\begin{align}
&\epsilon_{0}\nabla\left[\varepsilon(\mathbf{r})\nabla
\phi(\mathbf{r})\right]=q\left\{\rho_{e}(\mathbf{r})-\rho_{h}(\mathbf{r})-N_{dop}(\mathbf{r})\right\}.\label{eq:poiss} 
\end{align}
In \eqref{eq:ddcurr_el} and \eqref{eq:ddcurr_hl}, $\mu$ and $D$ denote the carrier mobility and diffusion constant, respectively. The terms on the right-hand side of Eqs.~\eqref{eq:conteq_el} and \eqref{eq:conteq_hl} describe the carrier generation and recombination rates. The generation rate $\mathcal{G}$ is related to the optical absorption via the imaginary part of the dielectric function $\varepsilon$ - here assumed to be isotropic - and the transverse part of the electrical field $\boldsymbol{\mathcal{E}}_{t}$ \cite{jackson}, 
\begin{align}
\mathcal{G}(\mathbf{r})=\int d\omega~\hbar^{-1}\eta_{gen}(\omega)\varepsilon_{0} \Im\mathrm{m}\mathbf{\varepsilon}(\mathbf{r},\omega)|\boldsymbol{\mathcal{E}}_{t}(\mathbf{r},\omega)|^2,
\end{align}
where $\eta_{gen}$ is the generation efficiency, which, here, is set to unity. 
For the recombination rate $\mathcal{R}$, only the fundamental radiative process is considered here. Conventionally, the radiative rate is related to the charge carrier densities as follows \cite{roosbroeck:54}: 
\begin{align}
\mathcal{R}(\mathbf{r})&=\mathcal{B}(\mathbf{r})\rho_{n}(\mathbf{r})\rho_{p}(\mathbf{r}),\label{eq:emrate}\\
\mathcal{B}(\mathbf{r})&=n_{i}^{-2}\int d\omega~\alpha(\mathbf{r},\omega)\phi_{bb}(\omega),\label{eq:vrs}
 \end{align}
 where $n_{i}$ denotes the intrinsic carrier density, \mbox{$\alpha\propto\Im\mathrm{m}~\varepsilon$} is the absorption coefficient and $\phi_{bb}(\omega)=\omega^2 n_{r}^2/\big(\pi^2c^2\big)\left\{\exp\left(\frac{\hbar\omega}{k_{B}T}\right)-1\right\}^{-1}$ is the angle-integrated black-body radiation flux, with refractive index $n_{r}$. 
 
For the solution of the above system of differential equations, boundary conditions need to be specified. The behavior of charge carriers at the contacts is described in terms of a surface recombination current,
\begin{align}
j_{s}^{B}=-q R^{B}_{s}=-q S_{s}\big(\rho_{s}-\rho_{s,eq}\big),\quad s=n,p,
\end{align}
where $S$ is the surface recombination velocity, which depends on the contact specifics such as passivation etc.   

\subsection{Quantum-kinetic device simulation: NEGF-Poisson}

On the quantum-kinetic level, the current and the rate term in the steady-state conservation law for charge carriers is formulated in terms of the charge carrier Green's function components $G^{R/A,\lessgtr}$ and scattering self-energies $\Sigma^{R/A,\lessgtr}$ \cite{kadanoff:62,keldysh:65,ae:prb_11},
\begin{align}
\nabla\cdot
&\mathbf{j}_{s}(\mathbf{r})
=\mp\frac{2e}{\hbar}\int\frac{dE}{2\pi}\int d\mathbf{r}'\Big[\Sigma^{R}_{s}
(\mathbf{r},\mathbf{r}',E)G_{s}^{<}(\mathbf{r}',\mathbf{r},E)
\nonumber\\&+\Sigma_{s}^{<}(\mathbf{r},\mathbf{r}',E)G_{s}^{A}(\mathbf{r}',\mathbf{r},E)
-G_{s}^{R}(\mathbf{r},\mathbf{r}',E)\Sigma_{s}^{<}(\mathbf{r}',\mathbf{r},E)\nonumber\\&-
G_{s}^{<}(\mathbf{r},\mathbf{r}',E)\Sigma_{s}^{A}(\mathbf{r}',\mathbf{r},E)\Big],\label{eq:conslaw_micro}\\
&\mathbf{j}_{s}(\mathbf{r})=\mp\lim_{\mathbf{r}'\rightarrow
\mathbf{r}}\frac{e\hbar}{m_{0}}\big(\nabla_{\mathbf{r}}-\nabla_{\mathbf{r}'}\big)
\int\frac{dE}{2\pi}G_{s}^{\lessgtr}(\mathbf{r},\mathbf{r}',E),
\end{align}
where the energy integration is over the band of carrier species $s=n,p$ and the upper (lower) sign is for electrons (holes).
Similarly, the charge carrier densities entering Poisson's equation \eqref{eq:poiss} are given by the GF
\begin{align}
&\rho_{s}(\mathbf{r})=\mp i\int
\frac{dE}{2\pi}G_{s}^{\lessgtr}(\mathbf{r},\mathbf{r},E).\label{eq:negfdens}
\end{align}
Unlike in the semiclassical case, the conservation law \eqref{eq:conslaw_micro} is not solved directly for the NEGF. Instead, the retarded and advanced GF components follow from the steady-state Dyson equations 
\begin{align}
 G^{R(A)}({\mathbf r_{1}},{\mathbf
 r}_{1'},E)&=G_{0}^{R(A)}({\mathbf r_{1}},{\mathbf
 r}_{1'},E)\nonumber\\ +&\int d^{3}r_{2}\int
 d^{3}r_{3}G_{0}^{R(A)}({\mathbf r}_{1},
 {\mathbf r}_{2},E)\nonumber\\\times&\Sigma^{R(A)}({\mathbf r}_{2},{\mathbf
 r}_{3},E) G^{R(A)}({\mathbf r}_{3},{\mathbf
 r}_{1'},E),\label{eq:dyson}
 \end{align}
 where $G_{0}$ denotes the solution of the non-interacting system. The correlation functions are obtained from the Keldysh equations,
 \begin{align}
 G^{\lessgtr}({\mathbf r}_{1},{\mathbf
 r}_{1'},E)=&\int d^{3}r_{2}\int d^{3}r_{3} 
 G^{R}({\mathbf
 r}_{1}, {\mathbf r}_{2},E)\nonumber\\
 &\times \Sigma^{\lessgtr}({\mathbf
 r}_{2},{\mathbf r}_{3},E) G^{A}({\mathbf
 r}_{3},{\mathbf r}_{1'},E).\label{eq:keldysh}
\end{align}
Equations \eqref{eq:dyson} and \eqref{eq:keldysh} are solved self-consistently together with the equations for the interaction part of the self-energies $\Sigma$, which depend on the carrier NEGF. In a second - or outer - self-consistency loop, the computation of the NEGF is self-consistently coupled to Poisson's equation via the density expressions \eqref{eq:negfdens}. 

For the self-energies, interactions of charge carriers with photons and phonons are considered. This enables the description of photogeneration, radiative recombination, and relaxation processes. The self-energy expressions associated with the generation via photon-mediated transitions between bands $a$ and $b$ read \cite{ae:prb89_14} 
\begin{align}
&\Sigma_{aa}^{\lessgtr}(\mathbf{r}_{1},\mathbf{r}_{2},E)=\Big(\frac{e}{m_{0}}\Big)^{2}\sum_{\mu\nu}\int
dE_{\gamma} A_{\mu}(\mathbf{r}_{1},E_{\gamma})
p_{ab}^{\mu}(\mathbf{r}_{1})\nonumber\\&\qquad\times G_{bb}^{\lessgtr}(\mathbf{r}_{1},\mathbf{r}_{2},E\mp
E_{\gamma})A_{\nu}^{*}(\mathbf{r}_{2},E_{\gamma})p_{ab}^{\nu*}(\mathbf{r}_{2}),\label{eq:se_coh}
\end{align}
where $\mathbf{p}$ contains the momentum matrix elements and the vector potential $\mathbf{A}$ is related to the transverse electric field via $\boldsymbol{\mathcal{E}}_{t}(\omega)=i\omega\mathbf{A}(\omega)$. The self-energy for the incoherent coupling to field fluctuations describing the spontaneous emission is given by the first self-consistent Born approximation (SCBA)
\begin{align}
&\Sigma^{e\gamma,\lessgtr}(\mathbf{r},\mathbf{r}',E)=i\hbar\mu_{0}\Big(\frac{e}{m_{0}}\Big)^2\sum_{\mu\nu}\lim_{\mathbf{r}''\rightarrow\mathbf{r}} \frac{1}{2}\left\{\hat{p}^{\mu}(\mathbf{r})-\hat{p}^{\mu}(\mathbf{r}'')\right\}\nonumber\\&\times p^{\nu}(\mathbf{r}')
\int\frac{dE'}{2\pi\hbar}\mathcal{D}^{\lessgtr}_{\mu\nu}(\mathbf{r}'',\mathbf{r'},E') G^{\lessgtr}(\mathbf{r},\mathbf{r'},E-E'),\label{eq:se_elphot_inc}
\end{align}
where $\hat{\mathbf{p}}=-i\hbar\nabla$ is the momentum operator. 
A similar SCBA expression is used for the description of the self-energies for electron-phonon interaction, with the photon propagator and the momentum operator in \eqref{eq:se_elphot_inc} replaced by the phonon propagator and the gradient of the electron-ion potential\cite{ae:jcel_11}. 

In the NEGF formalism, extraction and injection of charge carriers at the contacts is considered by boundary self-energies $\Sigma^B$. The exact form of these self-energies depends on the nature of the contact states. The retarded component encodes the contact-induced broadening and shifting of the device states, while the "$\lessgtr$"-components contain the information of the occupation of contact states according to the chemical potentials $\mu_{B}$ of the equilibrated electrodes.

\section{Model implementation\label{sec:modimp}}

\subsection{Semi-classical simulation}
For the semiclassical description of the ultrathin \mbox{$p$-$i$-$n$} device, the commercial semiconductor device simulator $ASA$ ($A$dvanced $S$emiconductor $A$nalysis, TU Delft)  is used. In the configuration employed here, the solver combines a \mbox{1-D} drift-diffusion-Poisson model for charge transport with a \mbox{1-D} transfer matrix method (TMM) for coherent wave optics \cite{yeh:77}. While it is common in macroscopic device simulation to use different band structure models for optical and electronic properties - fully dispersive absorption coefficient and refractive index from broadband experiments for optics versus single band effective mass models for transport - this is not applicable here for the sake of comparability with the quantum kinetic approach, where transport and optical transitions are described consistently on the basis of the same electronic structure. Thus, all the bulk properties will be described by means of a simple effective mass model for decoupled valence ($v$) and conduction ($c$) bands, assuming isotropic and parabolic dispersions $\varepsilon_{v}(\mathbf{k})=-\hbar^2k^2/2m_{v}^{*}$, $\varepsilon_{c}(\mathbf{k})=E_{g}+\hbar^2k^2/2m_{c}^{*}$ ($E_{g}$: energy gap), which sets the energy zero at the top of the valence band. The corresponding absorption coefficient is then given by
\begin{align}
\alpha_{\mathrm{bulk}}^{pb}(\hbar\omega)=&\frac{(2m_{r}^{*})^{\frac{3}{2}}}{2\pi\hbar^3\sqrt{\varepsilon_{b}}\varepsilon_{0}c_{0}\omega}\Big(\frac{e}{m_{0}} p_{cv}\Big)^2\sqrt{E-E_{g}}\Theta(E-E_{g}),\label{eq:abscoef_bulk}
\end{align} 
which provides the coupling of electronic and optical models via the extinction coefficient $\kappa(\omega)=\alpha\hbar c_{0}/(2\hbar\omega)$. In \eqref{eq:abscoef_bulk}, $m_{r}=(m_{c}^{-1}+m_{v}^{-1})^{-1}$ is the reduced effective mass, $\varepsilon_{b}$ is the background dielectric constant, $\varepsilon_{0}$ is the vacuum permittivity, $c_{0}$ is the speed of light in vacuum, and $p_{cv}$ is the interband momentum matrix element. The refractive index describing the light propagation in the TMM together with $\kappa$ is approximated by the constant value $n_{r}=\sqrt{\varepsilon_{b}}$. This absorption coefficient is also used to obtain the emission coefficient via the Van Roosbroeck-Shockley (VRS) relation \eqref{eq:vrs} in the calculation of the radiative recombination rate \eqref{eq:emrate}. The effective density of states required to relate the charge carrier densities to the quasi-Fermi levels (the quantity computed by $ASA$) is given by
\begin{align}
N_{c,v}=2\left(\frac{m^{*}_{c,v}k_{B}T}{2\pi\hbar^{2}}\right)^{\frac{3}{2}},
\end{align}
which also provides the intrinsic carrier density via $n_{i}^2\approx N_{c}N_{v}e^{-E_{g}/k_{B}T}$.
At the current stage, no microscopic model is used to parametrize the charge carrier mobilities $\mu_{c,v}$, but typical values from the literature are assumed. The contacts are described as ideally passivated and with full carrier selectivity via $S_{p}^{el}=S_{n}^{hl}=0$. The numerical values used for the material parameters are given in Tab.~\ref{tab:matpar_dd}. For the gold reflector, the optical data are taken from the SOPRA database \cite{sopra}.

\begin{table}[!t]
\renewcommand{\arraystretch}{1.3}
\caption{Material parameters used in the drift-diffusion simulation}
\label{tab:matpar_dd}
\centering
\begin{tabular}{|c||c|c|c|c|c|}
& $m^{*}/m_{0}$&$\mu$ (m$^2$/Vs)&$\varepsilon_{b}$&$E_{g}$ (eV)&$p_{cv}^2/m_{0}$ (eV)\\
\hline
electrons & 0.067&0.4&13.18&1.42&4.8\\
holes&0.1&0.04&&&
\end{tabular}
\end{table}

\subsection{Quantum-kinetic simulation}
The NEGF approach is implemented using the same combination of single-band effective mass models for electrons and holes and relying on the equivalence of the finite difference and the single-band tight-binding formalisms\cite{lake:97}. Under the assumption of periodicity in the transverse dimensions, Eq.~\eqref{eq:keldysh} and the differential form of Eq.~\eqref{eq:dyson} are discretized in real space in transport direction ($z$) and provide linear equations for the transverse Fourier components $G(\mathbf{k}_{\parallel},z,z',E)$. The numerical parameters for the resolution of position, transverse momentum, and energy are  $\Delta z$=5 $\mathring{A}$, $\Delta k_{\parallel}=0.002$ $\pi/d_{0}$, and $\Delta E$=4 meV, where $d_{0}$ is the lattice constant.

\begin{table}[!b]
\renewcommand{\arraystretch}{1.3}
\caption{Numerical parameters used in the NEGF simulation}
\label{tab:matpar_negf}
\centering
\begin{tabular}{|c|c|c|c|c|}
$\varepsilon_{\infty}$&$\hbar\Omega_{LO}$ (meV)&$D_{\textrm{ac}}$ [e/h] (eV)&$\rho$  (kg/m$^3$)&$c_{s}$ (m/s)\\
\hline
10.89&36&9/5&2329 &9040 
\end{tabular}
\end{table}

For the numerical evaluation of electron-phonon scattering, the self-energies for interaction of charge carriers with longitudinal polar optical (POP) and acoustic (AC) phonons are implemented, starting from the Fr\"ohlich Hamiltonian (POP) and the deformation potential formalism (AC), respectively\cite{lake:97}. The corresponding material parameters are given in Table~\ref{tab:matpar_negf}. 

The electron-photon interaction is treated as in \cite{ae:jpe_14}, with the classical vector potential $\mathbf{A}$ in \eqref{eq:se_coh} obtained from a quasi-1-D TMM (equivalent to that in $ASA$) and using in \eqref{eq:se_elphot_inc} the free field photon propagator \begin{align}
\mathcal{A}^{-1}\sum_{\mathbf{q}_{\parallel}}&\mathcal{D}^{>}_{\mu\nu,0}(\mathbf{q}_{\parallel},z,z',E_{\gamma})=-\frac{i
n_{0}E_{\gamma}}{3\pi\hbar c_{0}}\delta_{\mu\nu}
\end{align}
for a homogeneous isotropic optical medium ($\mathbf{q}_{\parallel}$ is the in-plane photon momentum). This corresponds to the assumptions of the semiclassical approach and neglects optical confinement and plasmonic effects from metal components on the emission \cite{niv:12}, for which a full solution of the photon GF would be required \cite{ae:oqel_14}. The extinction coefficient used in the TMM is again obtained from the absorption coefficient, which now reads
 \begin{align}
\alpha_{\mu}(\mathbf{q}_{\parallel},z,E_{\gamma})\approx&\frac{\hbar c_{0}}{2
n_{r}(\mathbf{q}_{\parallel},z,E_{\gamma})E_{\gamma}}\nonumber\\&\times\int dz'
\mathrm{Re}\Big[i\Pi_{\mu\mu}^{>}(\mathbf{q}_{\parallel},z',z,E_{\gamma})\Big],\label{eq:abscoef_negf}
\end{align}
in terms of the photon self-energy
\begin{align}
\Pi_{\mu\nu}^{>}(\mathbf{q}_{\parallel},z,z',E_{\gamma})=&-i\hbar\mu_{0}\Big(\frac{e}{m_{0}}\Big)^{2}p_{cv}^{\mu*}(z)p_{cv}^{\nu}(z')\nonumber\\&\times
\mathcal{P}_{cv}^{>}(\mathbf{q}_{\parallel},z,z',E_{\gamma})
\end{align}
based on the charge carrier GF via the interband polarization function
\begin{align}
 \mathcal{P}_{cv}^{>}(\mathbf{q}_{\parallel},z,z',E_{\gamma})=&\mathcal{A}^{-1}\sum_{\mathbf{k}_{\parallel}}\int
 \frac{dE}{2\pi\hbar}G_{cc}^{>}(\mathbf{k}_{\parallel},z,z',E)\nonumber\\&\times
 G_{vv}^{<}(\mathbf{k}_{\parallel} -\mathbf{q}_{\parallel},z',z,E-E_{\gamma}).\label{eq:polfun}
\end{align}
Unlike in the macroscopic approach leading to the bulk absorption coefficient \eqref{eq:abscoef_bulk}, it is essential to consider the non-locality of the electronic states in the derivation of \eqref{eq:abscoef_negf}, i.e., the off-diagonal matrix elements ($z\neq z'$) of the charge carrier GF in the interband polarization function \eqref{eq:polfun} \cite{ae:prb89_14}. For the situation considered here, 25 off-diagonals (corresponding to a nonlocality range of 12.5 nm) provide a satisfactory compromise between accuracy and computational cost.

The boundary self-energy terms are determined by matching to bulk Bloch states of semi-infinite flat band contact regions\cite{lake:97}. Perfect carrier selectivity is enforced by setting $\Sigma_{c}^{B}(z_{\mathrm{min}})=\Sigma_{v}^{B}(z_{\mathrm{max}})=0$. This prevents the leakage of carriers under optical and electronic injection \cite{cavassilas:15}.

\section{Numerical results and discussion\label{sec:resdisc}}

\begin{figure}[t]
\begin{center}
\includegraphics[width=0.45\textwidth]{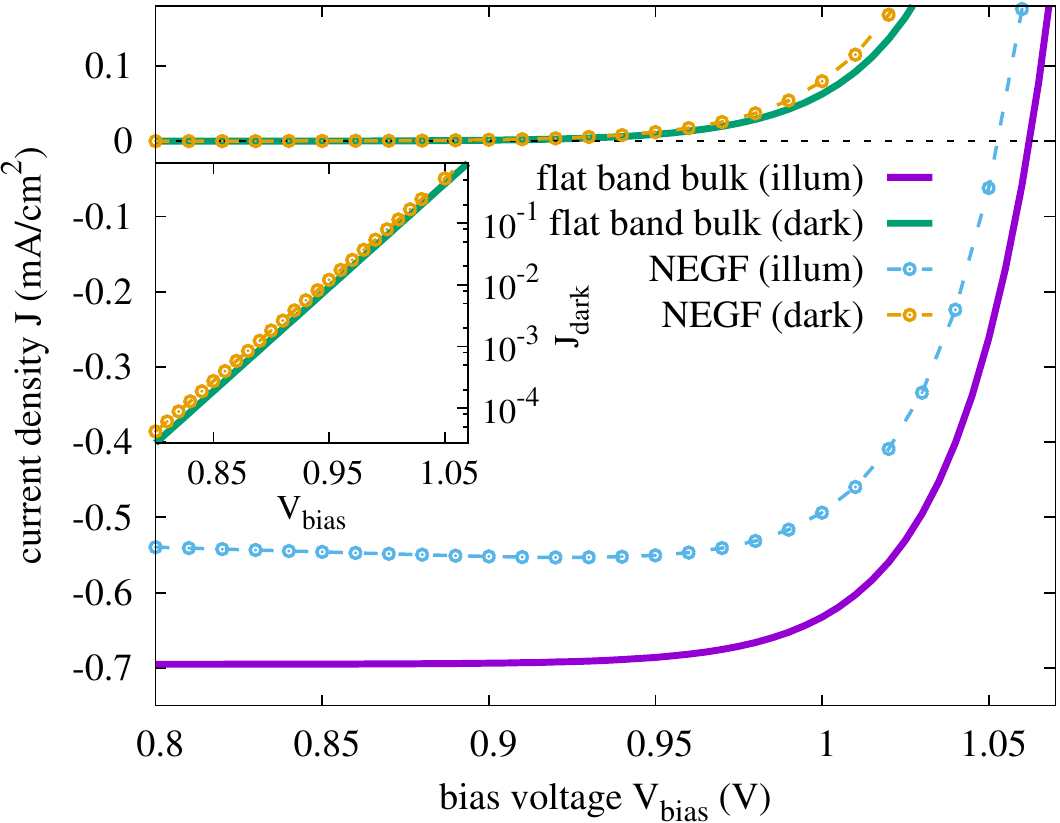}
\caption{Current-voltage characteristics of the ultrathin film solar cell device in the dark and under monochromatic illumination with photon energy $E_{\gamma}=1.44$ eV at an intensity of 0.1 W cm$^{-2}$. The semiclassical drift-diffusion-Poisson results for a flat-band bulk absorption coefficient are displayed by solid lines and the quantum-kinetic NEGF simulations by dashed lines with open symbols. The inset displays the dark $J-V$ curves in log scale. The characteristics exhibit significant qualitative and quantitative discrepancies between classical and quantum approaches both in the dark and under illumination.    \label{fig:JV_ASA_NEGF}}
\end{center}                     
\end{figure}

The room temperature ($T=300$ K) current-voltage characteristics of the slab diode are evaluated in the dark for bias voltages ranging from 0.8 to 1.07 V and in the same voltage range for monochromatic illumination at a photon energy $E_{\gamma}=1.44$ eV, which corresponds to generation close to the bulk band edge. The intensity is 0.1 W/cm$^2$. The maximum photocurrent that can be obtained from this illumination is  69.4 mA/cm$^2$. However, without an antireflection coating, the optical reflection loss of the system air-GaAs-Au is very high ($n_{Au}@1.44eV=0.2$, $\kappa_{Au}@1.44eV=5.6$). A small additional loss is due to parasitic absorption by the gold reflector. Hence, the current-voltage characteristics displayed in Fig. \ref{fig:JV_ASA_NEGF} exhibit much lower photocurrent. While the results from the semiclassical simulation for flat-band bulk material and those from the quantum-kinetic simulation are of the same order of magnitude, significant discrepancies, both of quantitative and qualitative nature, can be noted in the characteristics.

\begin{figure}[t]
\begin{center}
\includegraphics[width=0.45\textwidth]{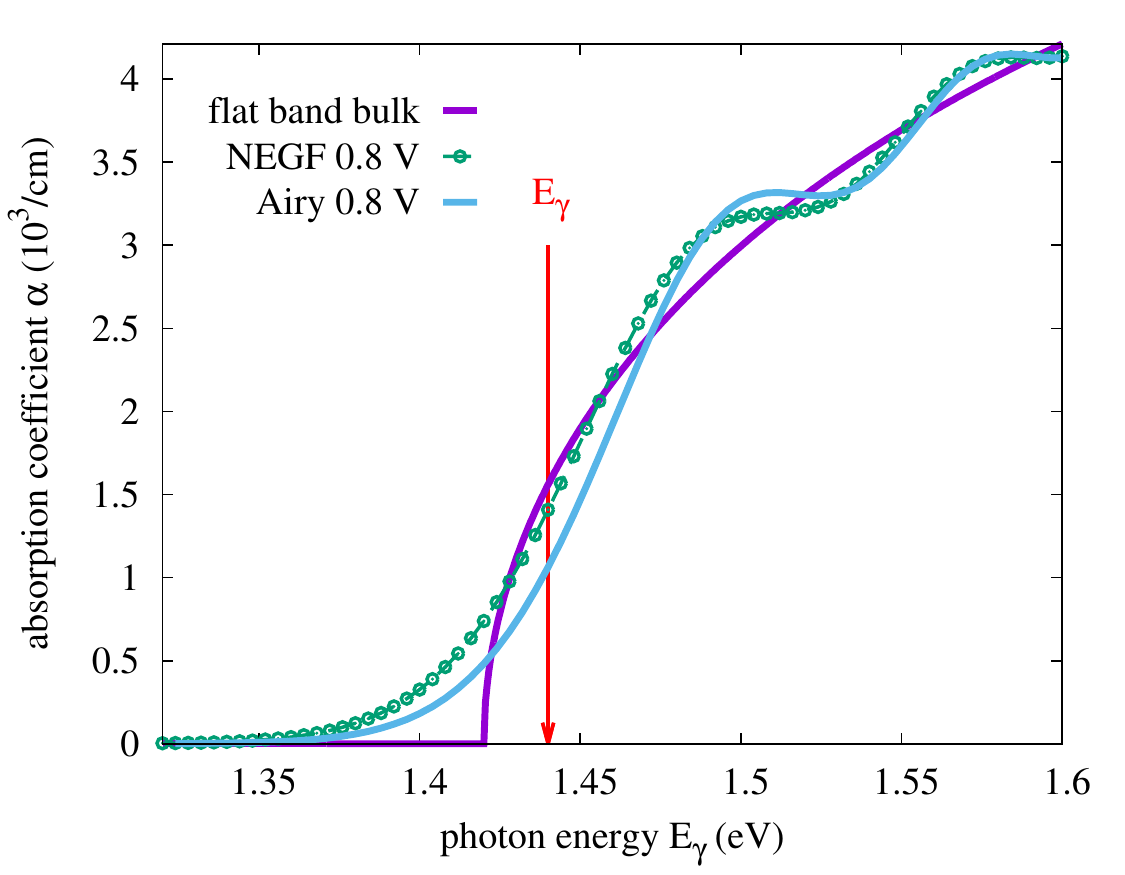}
\includegraphics[width=0.45\textwidth]{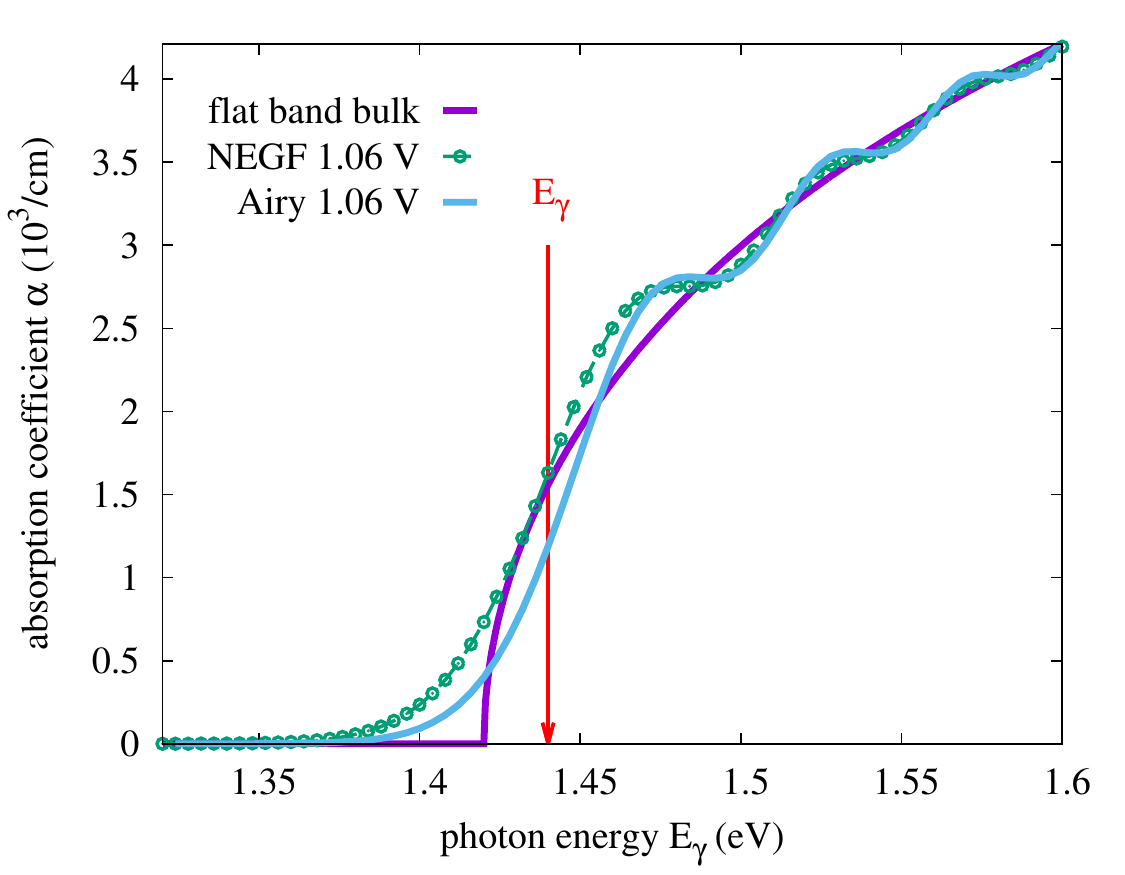}
\caption{Absorption coefficient for bulk material at flat-band conditions (dark solid line) compared with the NEGF absorption coefficient evaluated in the center of the intrinsic region at bias voltage $V_{\mathrm{bias}}=0.9/1.06$ V (symbols) and the Airy-function expression of electroabsorption for the central field values at these bias voltages (light solid line). The tailing and oscillatory behavior of the NEGF absorption and the resulting deviation from the bulk absorption is qualitatively captured by the Airy-function description of the field effects. Accordingly, the deviations are stronger for lower bias due to larger associated field strength. \label{fig:abscoef_bulk_NEGF_low_high}}
\end{center}
\end{figure}

At low bias voltage, the discrepancy is due to differences in the (spectral) absorptivity. Indeed, the comparison of the flat-band bulk absorption coefficient with the absorption coefficient in the center of the intrinsic region at 0.8 and 1.06 V as computed from the NEGF reveals substantial deviations over the whole spectral range considered (see Fig.~\ref{fig:abscoef_bulk_NEGF_low_high}). The appearance of sub-bandgap tails and oscillations at higher energies are well-known signatures of electroabsorption. Replacing the flat-band bulk absorption coefficient with the Airy-function-based expression for electroabsorption at the corresponding longitudinal field $\mathcal{E}_{z}$\cite{haug:04},
\begin{align}
\alpha_{\mathrm{bulk}}^{Ai}(E_{\gamma},\mathcal{E}_{z})=&~\frac{m_{r}^{*}f^{\frac{1}{3}}}{2\sqrt{\varepsilon}\varepsilon_{0}c_{0}\hbar E_{\gamma}}\left(\frac{e}{m_{0}}\right)^2 p_{cv}^2\nonumber\\&\times\Big(-\varepsilon
Ai^{2}(\varepsilon)+\left[Ai'(\varepsilon)\right]^2\Big),\label{eq:fieldabs}
\end{align}
with
\begin{align}
f=&~e\mathcal{E}_{z}\frac{2m_{r}^{*}}{2\pi\hbar^2},\quad
\varepsilon=\frac{2m_{r}^{*}(E_{g}-E_{\gamma})}{\hbar^2 f^{\frac{2}{3}}},
\end{align}
reproduces qualitatively the nonbulk-like features in the NEGF absorption coefficient. The electric field and thus the local absorption coefficient vary both with position within the device and with applied bias voltage. For the consideration of these effects in the semiclassical simulation, the local extinction coefficient for the TMM is determined from the   field-dependent absorption coefficient \eqref{eq:fieldabs} using the local field profile as computed for a given applied bias voltage in the dark, i.e., neglecting the effects of illumination.

Since the deviation from the flat-band bulk absorption stems primarily from the effect of the strong internal field on the (joint) density of states, it will also affect the emission properties. Indeed, comparison of the dark characteristics displayed in the inset of Fig.~\ref{fig:JV_ASA_NEGF} reveals a converging behavior concerning the ideality factor, but a substantial discrepancy in the dark saturation current, which is related to the emission coefficient. For a consistent description of absorption and emission, the  emission coefficient $\mathcal{B}$ needs to acquire a dependence on bias and position via the use of the field-dependent absorption coefficient \eqref{eq:fieldabs} in the VRS formula \eqref{eq:vrs}.

\begin{figure}
\begin{center}
\includegraphics[width=0.45\textwidth]{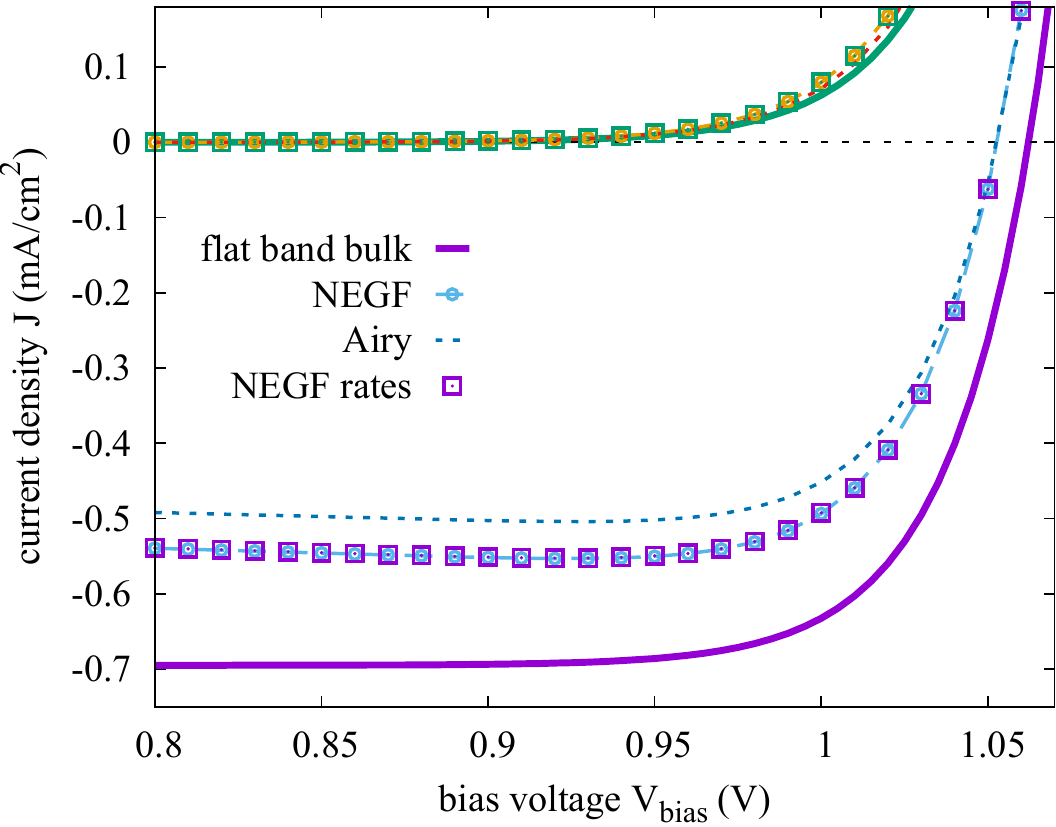}
\caption{Current-voltage characteristics including the semiclassical model with field-dependent absorption and emission coefficients, as well as with the full NEGF rates, in comparison with the semiclassical model with flat-band bulk absorption and the full NEGF model. Combination of the NEGF rates with the drift-diffusion transport reproduces the full NEGF result, which points at negligible quantum effects in the transport.\label{fig:JV_all}}
\end{center}            
\end{figure}

\begin{figure}[b]
\begin{center}
\includegraphics[width=0.5\textwidth]{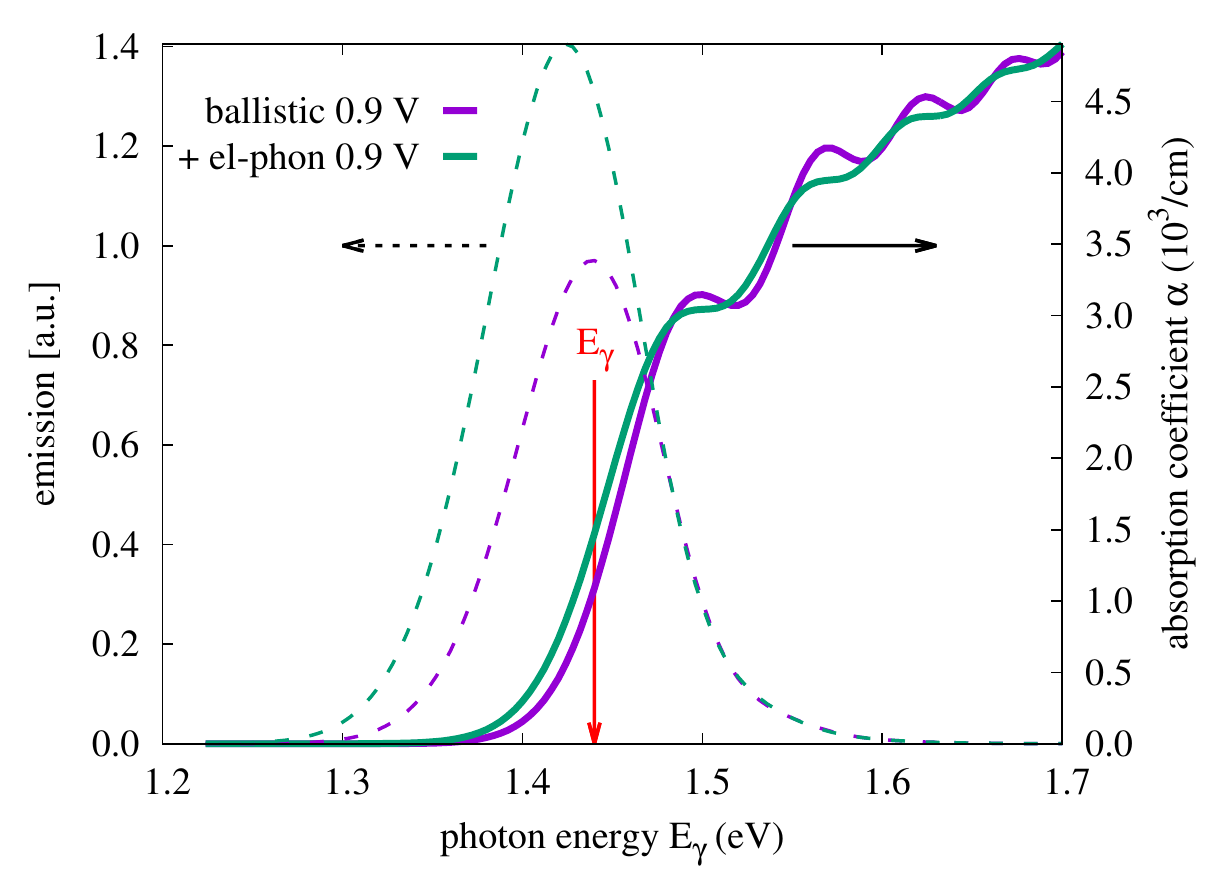}
\caption{Absorption coefficient and emission spectrum at $V_{\mathrm{bias}}=0.9$ V, as computed from the NEGF for a ballistic system and including electron-phonon interaction, respectively. The presence of inelastic scattering leads to lower absorption at the excitation energy and to larger emission, which corresponds exactly to the observed relation between Airy-function and NEGF characteristics.\label{fig:absem_ball_phon}}
\end{center}
\end{figure}

\begin{figure}[t]
\begin{center}
\includegraphics[width=0.5\textwidth]{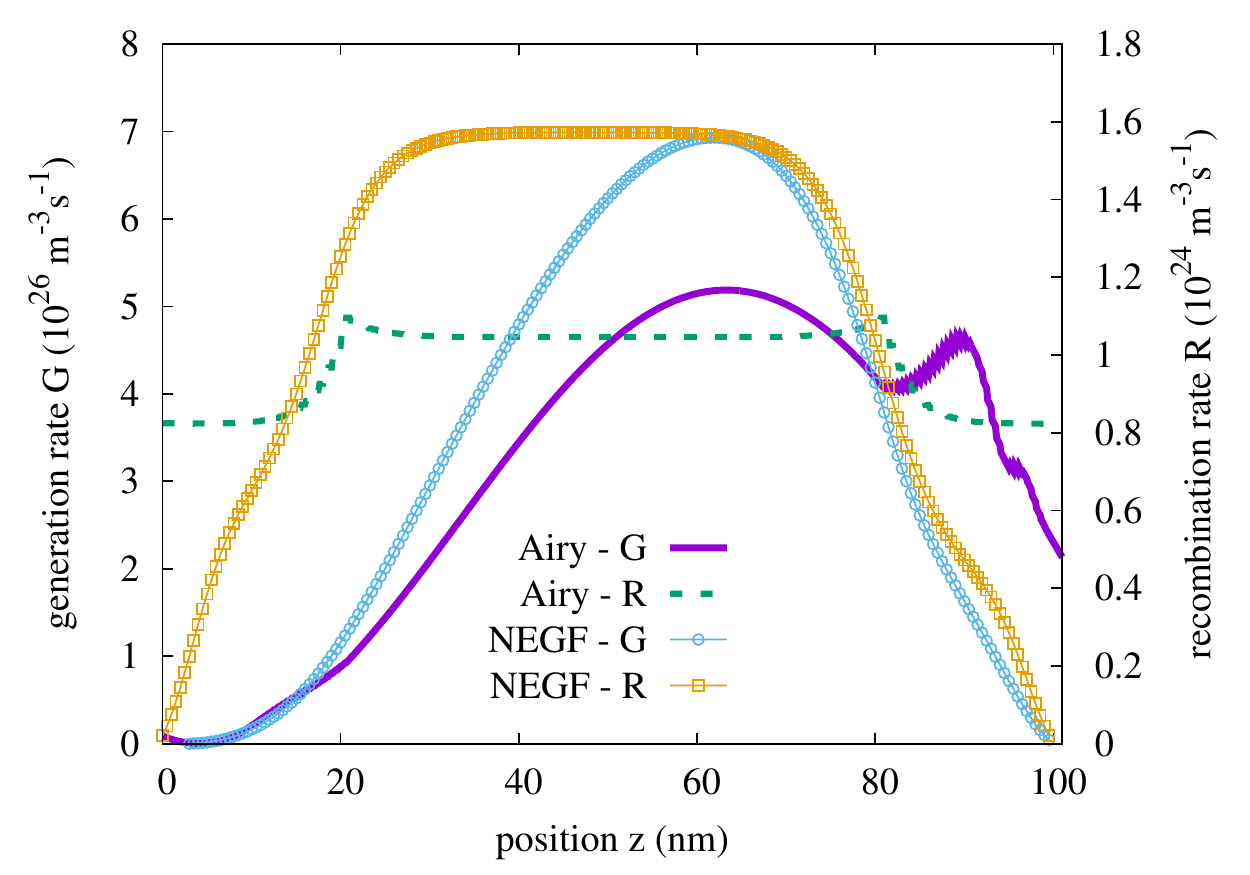}
\caption{Spatial variation of the volume generation and recombination rates at $V_{\mathrm{bias}}=0.9$ V for the semiclassical model with field-dependent absorption and emission coefficient (Airy, lines) and for the quantum-kinetic model (NEGF, symbols). While the average rates are similar, there are significant discrepancies in the spatial variation of generation and recombination, most notably close to the contacts.\label{fig:rates}}
\end{center}
                    
\end{figure}

As can be verified in Fig.~\ref{fig:JV_all}, consideration of the field effects provides a closer agreement of the semiclassical simulation with the quantum-kinetic result, both in the dark and under illumination. However, there is still some quantitative disagreement. One reason is the impact of electron-phonon interaction: As shown in Fig.~\ref{fig:absem_ball_phon}, the presence of inelastic electron-phonon scattering leads to lower absorption ($\rightarrow$ photocurrent) and larger emission ($\rightarrow$ dark current) as compared with the ballistic case, which corresponds exactly to the relation between the Airy-function and the NEGF characteristics in  Fig.~\ref{fig:JV_all}. 
The second source of discrepancies is the spatial variation of the generation and recombination rates within the device, especially close to the contacts (see Fig.~\ref{fig:rates}). This is due to the treatment of minority carrier contacts in the NEGF formalism, enforcing nodes in the charge carrier density at those contacts. To investigate the impact of deviations in the rate on the overall device characteristics, the NEGF rates for generation and recombination are directly used in the semiclassical balance equations \eqref{eq:conteq_el}-\eqref{eq:conteq_hl}. While $ASA$ is able to handle external generation files, the recombination rate needs to be engineered via the emission coefficient, using relation \eqref{eq:emrate}: The emission coefficient $\mathcal{B}$ is adjusted such that expression \eqref{eq:emrate} with the semiclassical charge carrier densities reproduces the NEGF emission rate. The result displayed in   Fig.~\ref{fig:JV_all} (open squares) reveals a perfect agreement with the full quantum kinetic simulation. This means that, for the situation under consideration, where transport is mediated by states that are still rather extended, the use of a semiclassical drift-diffusion model for charge carrier transport is valid. At the same time, a microscopic model with spatial resolution is required to capture the deviations from bulk rates induced by the strong internal fields. Indeed, because the density of states (DOS) is the property most affected by the field, a larger impact is to be expected on spectral quantities - such as absorption -  than on integral quantities like the effective DOS entering the transport model, since there, some of the field-induced spectral modifications are integrated out.

If the absorber thickness is reduced further, the increase in the strength of the built-in field at low bias voltage will result in an enhancement of the radiative recombination due to emission from field-induced tail states, as described in \cite{ae:jpe_14}, similar to tunneling-enhanced recombination via defects \cite{rau:99}. This feature is only captured by the NEGF model. On the other hand, the photocurrent generation level follows the size dependence of optical resonator modes \cite{niv:12}, which, in the quantum model, is only slightly modified, as compared with the classical model due to a moderate change in local absorption coefficients. Hence, the discrepancy between the uncorrected semiclassical and the quantum model is expected to become more pronounced with further thickness reduction. For a more general picture, additional losses due to finite surface recombination of minority carriers (leakage) and - in the ultrascaled limit - onset of band-to-band tunneling will need to be considered.   

\section{Conclusion\label{sec:sumconc}}

In this paper, it has been demonstrated that straight-forward application of semiclassical models to the simulation of ultrathin solar cells leads to significant discrepancies with respect to the more generally valid microscopic picture. This is most manifest in both spatial and bias dependence of photogeneration and radiative recombination rates due to the strong impact of the large built-in field on absorption and emission.

The results from the full quantum kinetic model can be recovered if the NEGF rates are used in the semiclassical transport equations. This means that transport is still well described by the drift-diffusion model, deviations from bulk DOS at the band edges being integrated out.

The picture is expected to change for the case where coherence plays a role in the transport, as in superlattice absorbers \cite{ae:nrl_11}, similarly to the situation found in THz quantum cascade lasers\cite{jirauschek:14}, or in the presence of barrier layers at the contacts\cite{cavassilas:15}.  There, at least partial breakdown of the validity of the semiclassical transport description is expected, necessitating the use of quantum-kinetic models, such as the one presented here.

Finally, one of the main hopes associated with ultrathin absorber architectures is the reduction of Shockley-Read-Hall (SRH) recombination due to fast extraction of photogenerated charge carriers by the strong built-in field \cite{steiner:13}. However, strong internal fields can also lead to enhanced SRH recombination due to field-assisted tunneling into defect states \cite{rau:99}. Extension of the present NEGF approach to the nonradiative regime \cite{ae:mrs_12} will, thus, be instrumental for a comprehensive assessment of field effects in ultrathin solar cell devices.

\section*{Acknowledgment}

The author would like to thank Bart Pieters for the introduction to $ASA$ and for helpful discussions.


\begin{IEEEbiography}[{\includegraphics[width=1in,height=1.25in,clip,keepaspectratio]{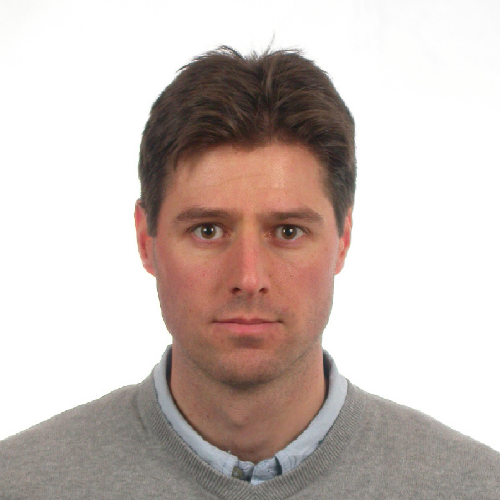}}]{Urs
Aeberhard} obtained his PhD in physics from ETH Z\"urich, Switzerland, in 2008 with a thesis on the quantum-kinetic theory of quantum-well solar cell
devices, carried out in the Condensed Matter Theory Group at Paul Scherrer Institute under the supervision of Dr. Rudolf Morf. From 2009-2012, he was a postdoctoral researcher at the Institute of Energy and Climate 
Research 5 - Photovoltaics at Forschungszentrum J\"ulich, Germany, where he now heads the activities on theory and multi-scale device simulation. The focus of his research is on the development of theory and numerical simulation approaches for advanced nanostructure-based solar cell devices.
\end{IEEEbiography}

\end{document}